\documentclass[aps,print,onecolumn,showpacs]{revtex4}
\usepackage{amsfonts}
\usepackage{amsmath}
\usepackage{amssymb}
\usepackage{graphicx}

\begin{document}

\title{Symmetry breaking of a matter-wave soliton in a double-well potential formed by spatially confined spin-orbit coupling}
\author{Zhi-Jiang Ye, Yi-Xi Chen, Yi-Yin Zheng, Xiong-Wei Chen, Bin Liu}
\email{binliu@fosu.edu.cn}
\affiliation{School of Physics and Optoelectronic Engineering, Foshan University,
Foshan 528000, China }
\pacs{03.75.Lm, 05.45.Yv}

\begin{abstract}
We consider the symmetry breaking of a matter-wave soliton formed by spinor Bose-Einstein condensates (BECs) illuminated by a two-spot laser beam. This laser beam
introduces spin-orbit (SO) coupling in the BECs such that the SO coupling produces an effect similar to a linear double-well potential (DWP). It is well known that symmetry breaking in a DWP is an important effect and has been discussed in many kinds of systems. However, it has not yet been discussed in a DWP formed by SO coupling. The objective of this work is to study the symmetry breaking of spinor BECs trapped by a DWP formed by SO coupling. We find that two kinds of symmetry breaking, displacement symmetry breaking and bimodal symmetry breaking, can be obtained in this model. The influence of the symmetry transition is systematically discussed by controlling the interaction strength of the BECs and the distance between the center of the two spots. Moreover, because SO coupling violates
Galilean invariance, the influence of symmetry breaking in the moving system is also addressed in this paper.
\end{abstract}

\maketitle

\section{Introduction}

Matter-wave solitons have become an interesting subject of research due to
their potential applications in various fields, such as atomic interferometry,
quantum information processing, and atomic lasers \cite{P.Meystre}. Experimental research on a self-trapping soliton in Bose-Einstein condensates
(BECs) began with the creation of a dark soliton, followed by bright soliton and
bright soliton trains \cite{PRL83_5198,Science287_97,Science296_1290,Nature417_150}.
Many studies have shown that a cold-atom BEC is an excellent system for
studying solitons \cite{PRA99_033625,Annals of
Physics2019,PRE98_012224,EPL122_36001,pla379_2193}. In particular, a stable bright soliton
has already been shown to improve the performance of a Mach-Zehnder
interferometer compared to regular BECs \cite{PRL113_013002}. Thus, the creation
of stable soliton has become a fascinating area of research.

Recently, the Gross-Pitaevskii equation (GPE) with the LHY correction term has been proven to be a good
mechanism to generate stable quantum droplets \cite{PRL119_050403,Daillie2016,Wachtler20162,Petrov2015,Cabrera2018,Chomaz2016,Schmitt2016,PRL116_215301,unstable-vort-DD,PRA94_033619,PRA97_011602,NJP_YL2017,PRA_YL2018,PRL_YVK2019,Ferioli2019,Staudinger2018,Cikojevic2018,Astrakharchik2018,Inguscio,PRA97_053623,PRA98_023630,PRA99_053602,PRA98_053835}%
, and the GP equation with long-range dipole-dipole interactions can also create
stable matter-wave solitons in BECs \cite{ZRF1,Perdi2005,Tikhonenkov2008,PRA78_043614,XChen2017}. With the help of spin-orbit (SO) coupling, absolutely stable (ground-state) and
metastable matter-wave solitons in 2D and 3D free space have been
reported \cite{PRA87_013614,PRE89_032920,PRL115_253902,Rep Prog Phys
78_026001,CGH2017,Yongping2016,PRA95_063613,WL1,WL2,Dw_pra97_063607}. Moreover, stable excited state solitons \cite{CSF111_62,CSF103_232,Rongxuan2018,SakaguchiFOP,Chunqing2018}, gap solitons \cite{PRL111_060402,Gaplocal2018}, and solitons with novel vortices \cite{Sakaguchi2016Vort,PRA95_013608,Xunda2016,Bingjin2017,Bingjin2018,Shimei2018} have been reported to be created by SO coupling \cite{WL3,Wpang2018}, and similar configurations have also been realized in an optical system \cite{YVK2015Opt,Sakaguchi2016Opt,Haohuang2019Opt,DaiND88_2629,DaiND87_1675}. However, previous studies on 2D and 3D solitons in  BECs with SO coupling tacitly assumed that the SO couplings were a homogeneous effect in the entire space. Recently, by using an external laser beam of a finite width, it was shown that one can implement spatially confined SO coupling, i.e., an SO coupling defect, in spinor BECs in 1D \cite{PRA90_063621} and 2D \cite{CNSNS73_481} space. It is interesting to find that solitons are trapped and caught by the SO coupling with a spatially confined modulation. These results imply that spatially confined SO coupling can act as a trapping potential to trap the BECs in space.

\begin{figure}[tph]
{\includegraphics[width=0.5\columnwidth]{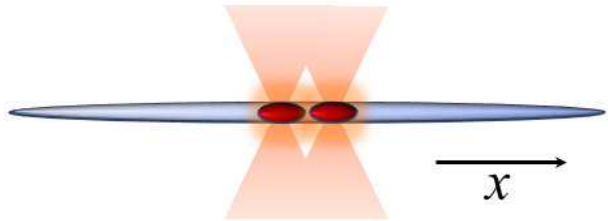}}
\caption{(Color online) Sketch of the system. The spinor BECs are illuminated by a two-spot SO coupling beam, which traps the BECs in a new type of DWP.}\label{sketch}
\end{figure}

If two of these beams, which form the SO coupling defect, are launched, an effective double-well potential (DWP) for the spinor BECs is created [see Fig.\ref{sketch}]. It is well known that one of the fundamental aspects of the nonlinear dynamics in the DWP is spontaneous
symmetry breaking (SSB), in which a symmetric state breaks its symmetry to a favorable asymmetric state. The concept of SSB in nonlinear systems was introduced by J. C. Eilbeck \cite{PhysicaD16_318}. Its manifestations can be found in a variety of settings, including classical and quantum mechanics, dual-core optical waveguides and Bragg gratings, nonlinear discrete systems, nonlinear optics and other physical systems \cite{PhysicaD16_318,PRB68_035325,PRE76_066606,PRA84_053618}. In particular, SSB effects were studied in detail in BECs \cite{PRA59_1457,PRA64_061603,Nature443_312,PRA79_013626,PRA81_053630,PRA87_013604,FrontPhys12_124206,Yongyao2012,JPB48_045301}. Applications of this effect, such as the design of power-switch devices based on soliton light propagation in fibers, were proposed \cite{book_ssb}. In BECs, the SSB of matter-wave solitons in a DWP has been considered in many configurations but not for matter-wave solitons in a DWP formed by spatially confined SO coupling. Hence, the mechanism of SSB in a DWP formed by SO coupling remains unclear.

The objective of this work is to study the SSB of matter-wave solitons in a DWP formed by two spatially confined SO couplings in a 1D system. A sketch of the configuration of this system is shown in Fig. \ref{sketch}: spinor BECs with an attractive interaction are trapped in a horizontal cigar-shaped potential well, and two external laser illuminations, which form the spatially confined SO coupling, trap the BECs in their illumination area. We study the mechanism of SSB in such a DWP and find two types of symmetry breaking, displacement symmetry breaking and bimodal symmetry breaking. The mechanism of these two kinds of SSB are systematically discussed in this paper. Moreover, because SO coupling violates Galilean invariance, discussing the SSB in the moving system is a nontrivial issue, which is also discussed in detail in this paper. The rest of the paper is structured as follows. The model is introduced in Sec. II. Basic numerical results for the SSB of a matter-wave soliton in the quiescent and moving reference frames are reported in Sec. III. The paper is concluded in Sec. V.

\section{The model}

We consider binary BECs with a pseudo-spinor wave function $\Psi
_{+}\left( x,t\right) $, $\Psi _{-}\left( x,t\right) $. The mean-field model
of the system is based on the Lagrangian%
\begin{eqnarray}
L &=&\int \mathcal{L}dx,  \notag \\
\mathcal{L} &=&-\frac{i}{2}\left( \Psi _{+}^{\ast }\frac{\partial \Psi _{+}}{%
\partial t}+\Psi _{-}^{\ast }\frac{\partial \Psi _{-}}{\partial t}%
-c.c.\right) +\frac{1}{2}\left( \left\vert \partial _{x}\Psi _{+}\right\vert
^{2}+\left\vert \partial _{x}\Psi _{-}\right\vert ^{2}\right)   \notag \\
&&-\frac{g}{2}\left( \left\vert \Psi _{+}\right\vert ^{4}+\left\vert \Psi
_{-}\right\vert ^{4}\right) -g\gamma \left\vert \Psi _{+}\right\vert
^{2}\left\vert \Psi _{-}\right\vert ^{2}+\frac{\lambda \left( x\right) }{2}%
\left\{ \left( \Psi _{+}^{\ast }\frac{\partial \Psi _{-}}{\partial x}-\Psi
_{-}^{\ast }\frac{\partial \Psi _{+}}{\partial x}\right) +c.c.\right\} ,
\label{lagrangian}
\end{eqnarray}%
where $c.c.$ stands for a complex conjugate, $\gamma $ is the relative
strength of the cross-attraction, and the strength of the self-attraction is
normalized to $1$. $g$ is the total nonlinear strength.
The SO coupling of the Rashba type, which has a double-well structure, is
\begin{eqnarray}
\lambda \left( x\right) =\lambda
_{0}\exp \left[-\left( x-x_{0}\right) ^{2}/D^{2}\right]+\lambda_{0}\exp\left[-\left( x+x_{0}\right)
^{2}/D^{2}\right], \label{DWP}
\end{eqnarray}where $\lambda _{0}\equiv 1$ is the normalized peak strength of the SO coupling, $x_{0}$ denotes the half distance between the two wells, and $D$ is the width of the well, i.e., the width of the spots of the illumination.

The Gross-Pitaevskii equation (GPE) is derived from Lagrangian (\ref{lagrangian}) using the Euler-Lagrange equations as follows:
\begin{eqnarray}
i\frac{\partial \Psi _{+}}{\partial t} &=&-\frac{1}{2}\frac{\partial
^{2}\Psi _{+}}{\partial x^{2}}-g\left( \left\vert \Psi _{+}\right\vert
^{2}+\gamma \left\vert \Psi _{-}\right\vert ^{2}\right) \Psi _{+}+\lambda
\left( x\right) \frac{\partial \Psi _{-}}{\partial x}+\frac{\lambda
_{x}\left( x\right) }{2}\Psi _{-},  \notag \\
i\frac{\partial \Psi _{-}}{\partial t} &=&-\frac{1}{2}\frac{\partial
^{2}\Psi _{-}}{\partial x^{2}}-g\left( \left\vert \Psi _{-}\right\vert
^{2}+\gamma \left\vert \Psi _{+}\right\vert ^{2}\right) \Psi _{-}-\lambda
\left( x\right) \frac{\partial \Psi _{+}}{\partial x}-\frac{\lambda
_{x}\left( x\right) }{2}\Psi _{+}.  \label{Model}
\end{eqnarray}

Stationary solutions to Eqs. (\ref{Model}) with a chemical potential $\mu $
are sought as%
\begin{equation}
\Psi _{\pm }\left( x,t\right) =\phi _{\pm }\left( x\right) e^{-i\mu t},
\label{ansatz}
\end{equation}%
where the functions $\phi _{\pm }\left( x\right) $ satisfy the equations
\begin{eqnarray}
\mu \phi _{+} &=&-\frac{1}{2}\frac{\partial ^{2}\phi _{+}}{\partial x^{2}}%
-g\left( \left\vert \phi _{+}\right\vert ^{2}+\gamma \left\vert \phi
_{-}\right\vert ^{2}\right) \phi _{+}+\lambda \left( x\right) \frac{\partial
\phi _{-}}{\partial x}+\frac{\lambda _{x}\left( x\right) }{2}\phi _{-},
\notag \\
\mu \phi _{-} &=&-\frac{1}{2}\frac{\partial ^{2}\phi _{-}}{\partial x^{2}}%
-g\left( \left\vert \phi _{-}\right\vert ^{2}+\gamma \left\vert \phi
_{+}\right\vert ^{2}\right) \phi _{-}-\lambda \left( x\right) \frac{\partial
\phi _{+}}{\partial x}-\frac{\lambda _{x}\left( x\right) }{2}\phi _{+},
\label{model_phi}
\end{eqnarray}%
which can be derived from their own Lagrangian density:%
\begin{eqnarray}
\mathcal{L}_{stat} &=&-\mu \left( \left\vert \phi _{+}\right\vert ^{2}+\left\vert \phi
_{-}\right\vert ^{2}\right) +\frac{1}{2}\left( \left\vert \partial _{x}\phi
_{+}\right\vert ^{2}+\left\vert \partial _{x}\phi _{-}\right\vert ^{2}\right)
\notag \\
&&-\frac{1}{2}g\left( \left\vert \phi _{+}\right\vert ^{4}+\left\vert \phi
_{-}\right\vert ^{4}\right) -\gamma \left\vert \phi _{+}\right\vert
^{2}\left\vert \phi _{-}\right\vert ^{2}
+\frac{\lambda \left( x\right) }{2}\left\{ \left( \phi _{+}^{\ast }\frac{%
\partial \phi _{-}}{\partial x}-\phi _{-}^{\ast }\frac{\partial \phi _{+}}{%
\partial x}\right) +c.c.\right\}  \label{L_stat}
\end{eqnarray}%

The total norm of ansatz (\ref{ansatz}) is
\begin{eqnarray}
N=\int n(x)dx=\int \left( \left\vert \phi
_{+}\right\vert ^{2}+\left\vert \phi _{-}\right\vert ^{2}\right) dx \label{Norm}
\end{eqnarray}
where $n(x)=|\phi_{+}(x)|^{2}+|\phi_{-}(x)|^{2}$ is the total density pattern soliton. The solution of the matter-wave soliton of Eq. (\ref{model_phi}) is solved numerically by means of the imaginary-time-integration method \cite{ITP1,ITP2}. In addition, the stabilities are verified by direct simulations. To study the mechanism of SSB in such a DWP clearly, we will apply the normalized condition to the total norm of the soliton and fix the value of $D$ in the numerical simulations. Hence, the free control parameters for the system are the nonlinear strength $g$, the strength of the cross-interaction $\gamma$, and the distance between the two wells, $x_{0}$. Here, larger values of $g$ and smaller values of $x_{0}$ correspond to a stronger attractive nonlinearity and coupling between the two wells, respectively.

\section{SYMMETRIC AND ASYMMETRIC SOLUTIONS}

\subsection{Quiescent reference frame}

\begin{figure}[tph]
{\includegraphics[width=0.75\columnwidth]{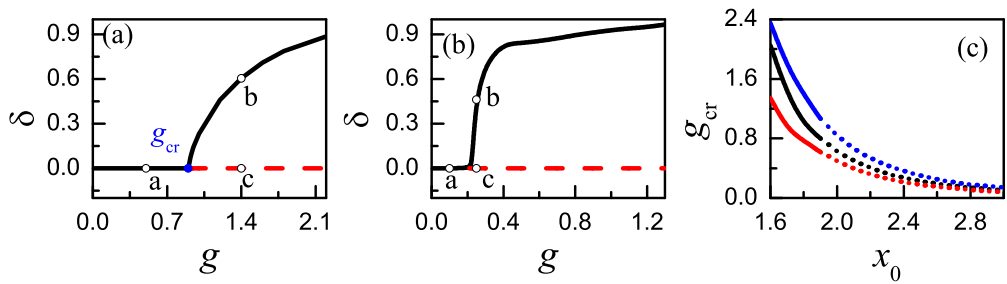}}
\caption{(Color online) Bifurcation diagrams for the symmetric and asymmetric
solutions, in the plane of ($g$, $\protect\delta $), as found from the numerical
solution of Eq. (\protect\ref{model_phi}) at different values of $x_{0}$.
(a) $x_{0}=1.85$; (b) $x_{0}=2.5$. (c) The blue, black, and red curves show the
displacement supercritical points for $\protect\gamma =0$, $1$, and $%
2 $, respectively. The blue, black, and red dashed curves show the bimodal
supercritical points for $\protect\gamma =0$, $1$, and $2$,
respectively.} \label{bifuraction}
\end{figure}

\begin{figure}[tph]
{\includegraphics[width=0.75\columnwidth]{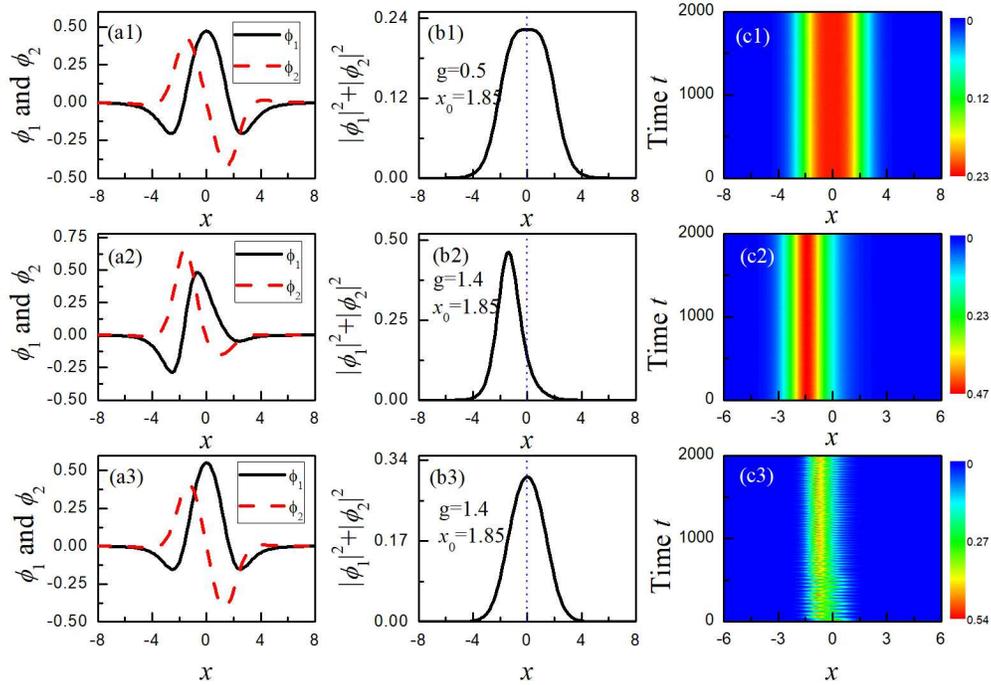}}
\caption{(Color online) The profiles of the $\protect\phi _{1}(x)$ and $\protect\phi %
_{2}(x)$ components of the solutions are shown by the solid black and dashed red
curves, respectively, for $g=0.5$, $1.4$, and $1.4$,
severally in panels (a1)-(a3). These example solutions correspond to the
points a, b, and c marked in Fig. 2(a). The densities, i.e., $n(x)$, are shown in panels
(b1)-(b3). The direct simulations of the perturbed evolution of $\left\vert
\protect\phi \right\vert ^{2}$ are shown in panels (c1)-(c2). Here, the perturbation is 3\% random noise, and the other parameters are $\protect\gamma %
=1 $, $D=2$ and $x_{0}=1.85$.} \label{displacement}
\end{figure}

\begin{figure}[tph]
{\includegraphics[width=0.75\columnwidth]{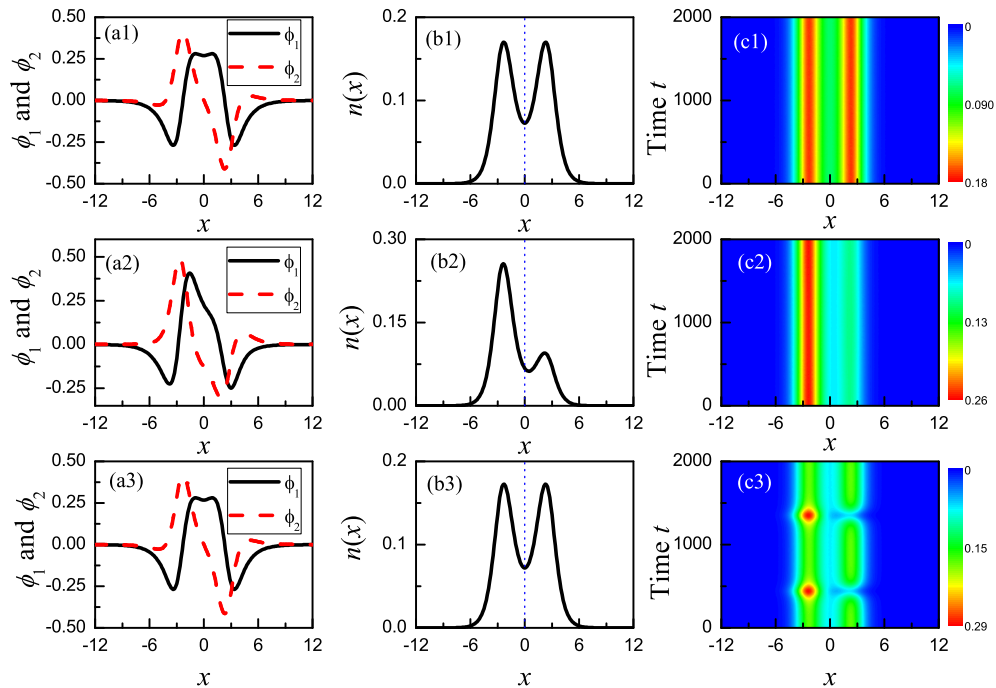}}
\caption{(Color online) The profiles of the $\protect\phi _{1}(x)$ and $\protect\phi %
_{2}(x)$ components of the solutions are shown by the solid black and dashed red
curves, respectively, for $g=0.1$, $0.25$, and $0.25$ in panels (a1)-(a3). These example solutions correspond to the points a, b, and c marked in Fig. 2(b). The densities, i.e., $n(x)$, are shown in panels
(b1)-(b3). The direct simulations of the perturbed evolution of $\left\vert
\protect\phi \right\vert ^{2}$ are shown in panels (c1)-(c2). Here, the
perturbation is 3\% random noise, and the other parameters are $\protect\gamma %
=1 $, $D=2$ and $x_{0}=2.5$.} \label{bimodal}
\end{figure}

To identify the mechanism of the SSB precisely, we define the asymmetric character as follows:
\begin{equation}
\delta =\left\vert \frac{n^{\max}_{L}-n^{\max}_{R}}{n^{\max}_{L}+n^{\max}_{R}}%
\right\vert ,  \label{delta}
\end{equation}%
where $n^{\max}_{L}$ and $n^{\max}_{R}$ are the peak values (i.e., maxima) of the total density pattern in the regions of $x\in(-\infty,0)$ and $(0,\infty)$, respectively. If $\delta=0$, the soliton has a symmetric density profile; otherwise, the soliton is asymmetric. Hence, the mechanism of the SSB can be characterized by a bifurcation diagram of $\delta$ by varying $g$ or $x_{0}$. The numerical simulations find that increasing the values of $g$ or reducing the values of $x_{0}$ can lead to SSB of the soliton, which means that a stronger attractive nonlinearity and weaker coupling between the wells can easily lead to the SSB. The numerical results demonstrate that the current DWP can feature a similar property as in the usual DWPs, which is $\sim-\lambda(x)$. Two different kinds of SSB, displacement SSB and bimodal SSB, are found in this system by increasing $g$ or reducing $x_{0}$. The former type of SSB (i.e., displacement SSB) breaks the symmetry of single-peak solitons by shifting their center of mass from the origin of coordinates (i.e., $x=0$), while the latter type of SSB (i.e., bimodal SSB) breaks the symmetry of a double-peak soliton by breaking the balance of the peak values. By increasing the values of $g$, displacement SSB tends to occur with a smaller value of $x_{0}$, whereas bimodal SSB occurs with a larger value of $x_{0}$. Figs. \ref{bifuraction}a and \ref{bifuraction}b show the bifurcation diagrams, which are displayed by $\delta$ versus $g$, for the displacement SSB and the bimodal SSB, respectively. The figures show that these two kinds of SSB are both supercritical, in which the asymmetric branches ($\delta\neq0$) emerge as the stable branch ($\delta=0$) and immediately go in the forward direction. Hence, these two types of SSB are tantamount to a phase transition of the second kind. Figs. \ref{displacement} and \ref{bimodal} display typical examples of stable and unstable solitons for the two kinds of SSB, which are selected from the bifurcation diagrams in Figs. \ref{bifuraction}a and \ref{bifuraction}b (see the points `a', `b', and `c' in these two panels).

In Figs. \ref{displacement}(a1), (a2) and (a3), one can see the amplitudes of the stable symmetric, asymmetric and unstable
symmetric solitons, respectively, which correspond to points `a', `b', and `c' in Fig. \ref{bifuraction}(a). These soliton solutions are complex functions that are specific to SO coupling (here, the imaginary parts is small). Their total density patterns, i.e., $n(x)$, which characterize their overall symmetry properties, are shown in Figs. \ref{displacement}(b1)-(b3). The direct simulations of the perturbed evolution of the soliton, which identify the stability and are shown by the evolution of the total density pattern, are displayed in panels \ref{displacement}(c1)-(c3). The reason why the phenomenon of displacement SSB tends to occur with smaller values $x_{0}$ can be explained as follows. The bottom of the DWP, i.e., $-\lambda(x)$, has a `W' profile, which is constructed by a local maximum (at $x=0$) and two adjacent minima (at $x=\pm x_{0}$). The difference between the center maximum and the two adjacent minima is denoted by the magnitude of $x_{0}$. When $x_{0}$ is small, the difference is also small and is not enough to modulate the density pattern of the soliton. Hence, the soliton can maintain a single-peak profile and rest at the center when the nonlinearity is not obvious. However, if the nonlinearity is enhanced, the ground state is created in the adjacent minima; hence, the soliton shifts its center of mass away from the coordinate origin.

Fig. \ref{bimodal} shows a similar figure construction as in Fig. \ref{displacement} for the case of bimodal SSB, which corresponds to points `a', `b', and `c' in Fig. \ref{bifuraction}(b). The reason why the phenomenon of bimodal SSB tends to occur with a large value of $x_{0}$ can be explained by similar reasons as follows. When $x_{0}$ is large, the difference between the local maximum (at $x=0$) and the adjacent minima (at $x=\pm x_{0}$) becomes large. If this difference is large enough to modulate the density profile of the soliton, a soliton with a balanced double-peak structure is created when the nonlinearity is small. Then, if the nonlinearity is enhanced, SSB occurs and transforms the balanced double-peak structure into an imbalanced one.

The bifurcation point, $g_{\mathrm{cr}}$ as a function of $x_{0}$ for different values of $\gamma$, is displayed in Fig. \ref{bifuraction}c. A smaller value of $g_{\mathrm{cr}}$ implies that SSB is easier to induce. As expected, the figure shows that $g_{\mathrm{cr}}$ decreases as $x_{0}$ increases, which means that reducing the coupling between the two wells can easily induce SSB. The dividing points between displacement SSB and bimodal SSB are labeled in the curve of $g_{\mathrm{cr}}(x_{0})$ by the junction between the solid and the dash curves. The figure also shows that the magnitude of $g_{\mathrm{cr}}$ is also strongly influenced by the value of $\gamma$. Since the larger value of $\gamma$ increases the effect of the total nonlinearity, the curve of $g_{\mathrm{cr}}(x_{0})$ with larger values of $\gamma$ is lower than the curve with smaller values of $\gamma$. However, the figure shows that the dividing point between the two types of SSB does not show an obvious relationship with $\gamma$.

\subsection{Moving reference frame}

\begin{figure}[tph]
{\includegraphics[width=0.5\columnwidth]{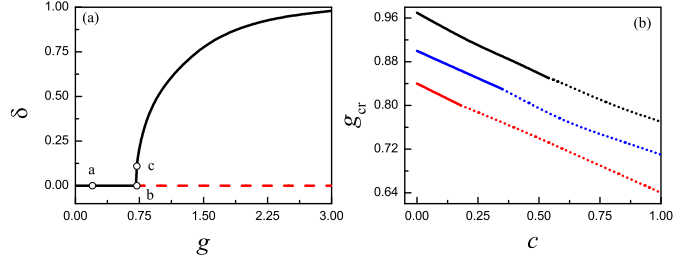}}
\caption{(Color online) (a) Bifurcation diagrams for the symmetric and
asymmetric solutions, in the plane of ($g$, $\protect\delta $), as found from the numerical solution of Eq. (\ref{move_model}) with $c=1$. (b) The black, blue, and red curves show
the displacement supercritical points for $x_{0}=1.825$, $1.85$, and $%
1.875$, respectively. The blue, black, and red dashed curves show the
bimodal supercritical points for $x_{0}=1.825$, $1.85$, and $1.875$,
respectively.}
\label{movingbifuraction}
\end{figure}

\begin{figure}[tph]
{\includegraphics[width=0.75\columnwidth]{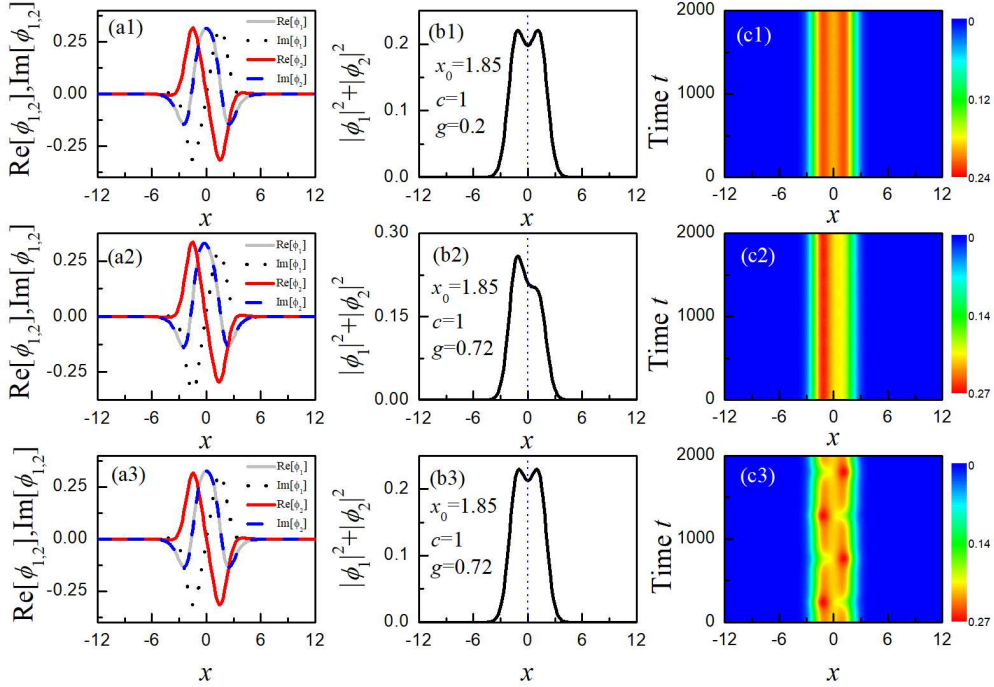}}
\caption{(Color online) The real and imaginary parts of the $\protect\phi _{1}(x)$
and $\protect\phi _{2}(x)$ components of the solutions are shown by the solid
gray, dotted black, solid red and dashed blue curves, respectively, for $g=0.2$, $0.72$, and $0.72$ in panels
(a1)-(a3). These example solutions correspond to the points a, b, and c marked in Fig. 5(a). The densities, i.e., $n(x)$, are shown in panels (b1)-(b3). The direct
simulations of the perturbed evolution of $\left\vert \protect\phi %
\right\vert ^{2}$ are shown in panels (c1)-(c2). Here, the perturbation is 3\%
random noise, and the other parameters are $\protect\gamma =1$, $D=2$, $%
x_{0}=1.85$ and $c=1$.}
\label{moving}
\end{figure}
\

Unlike the usual DWP, the system in SO coupling does not obey Galilean invariance. Hence, the studies of SSB in a DWP formed by SO coupling is a nontrivial issue. Here, we assume the stable transfer of solitons by the moving SOC profile, which corresponds to Eq. (\ref{DWP}) with
\begin{eqnarray}
\lambda(x)\rightarrow\lambda(x'),
\end{eqnarray}
where $x'=x-ct$. The purpose of this subsection is to determine the influence of $c\neq0$ on the SSB. For convenience, we will fix $\gamma=1$ in this subsection.

To address this issue, Eq. (\ref{Model}) is rewritten in terms of the moving coordinate with the transformed wave
function, $\Psi_{\pm}(x,t)=\Psi_{\pm}(x',t)$, as
\begin{eqnarray}
i\frac{\partial \Psi _{+}}{\partial t} &=&-\frac{1}{2}\frac{\partial
^{2}\Psi _{+}}{\partial x^{\prime 2}}+ic{\frac{\partial \Psi _{+}}{\partial
x^{\prime }}}-g\left( \left\vert \Psi _{+}\right\vert ^{2}+\gamma \left\vert
\Psi _{-}\right\vert ^{2}\right) \Psi _{+}+\lambda \left( x^{\prime }\right)
\frac{\partial \Psi _{-}}{\partial x}+\frac{\lambda _{x^{\prime }}\left(
x^{\prime }\right) }{2}\Psi _{-},  \notag \\
i\frac{\partial \Psi _{-}}{\partial t} &=&-\frac{1}{2}\frac{\partial
^{2}\Psi _{-}}{\partial x^{\prime 2}}+ic{\frac{\partial \Psi _{-}}{\partial
x^{\prime }}}-g\left( \left\vert \Psi _{-}\right\vert ^{2}+\gamma \left\vert
\Psi _{+}\right\vert ^{2}\right) \Psi _{-}-\lambda \left( x^{\prime }\right)
\frac{\partial \Psi _{+}}{\partial x}-\frac{\lambda _{x^{\prime }}\left(
x^{\prime }\right) }{2}\Psi _{+}.  \label{move_model1}
\end{eqnarray}%
Further, by applying the transformation
\begin{equation*}
\Psi _{\pm }(x^{\prime },t)=\Psi _{\pm }^{\prime }(x^{\prime },t)\exp
[icx^{\prime 2}/2],
\end{equation*}%
Eq. (\ref{move_model1}) can be transformed to
\begin{eqnarray}
i\frac{\partial \Psi _{+}^{\prime }}{\partial t} &=&-\frac{1}{2}\frac{%
\partial ^{2}\Psi _{+}^{\prime }}{\partial x^{\prime 2}}-g\left( \left\vert
\Psi _{+}^{\prime }\right\vert ^{2}+\gamma \left\vert \Psi _{-}^{\prime
}\right\vert ^{2}\right) \Psi _{+}^{\prime }+\lambda \left( x^{\prime
}\right) \frac{\partial \Psi _{-}^{\prime }}{\partial x^{\prime }}+\frac{%
\lambda _{x^{\prime }}\left( x^{\prime }\right) }{2}\Psi _{-}-ic\lambda
\left( x^{\prime }\right) \Psi _{-}^{\prime },  \notag \\
i\frac{\partial \Psi _{-}^{\prime }}{\partial t} &=&-\frac{1}{2}\frac{%
\partial ^{2}\Psi _{-}^{\prime }}{\partial x^{\prime 2}}-g\left( \left\vert
\Psi _{-}^{\prime }\right\vert ^{2}+\gamma \left\vert \Psi _{+}^{\prime
}\right\vert ^{2}\right) \Psi _{-}^{\prime }-\lambda \left( x^{\prime
}\right) \frac{\partial \Psi _{+}^{\prime }}{\partial x^{\prime }}-\frac{%
\lambda _{x^{\prime }}\left( x^{\prime }\right) }{2}\Psi _{+}^{\prime
}+ic\lambda \left( x^{\prime }\right) \Psi _{+}^{\prime }.
\label{move_model}
\end{eqnarray}%
In usual DWP systems, which are realized in homogeneous space, the transformation from Eq. (\ref{move_model1}) to (\ref{move_model}) makes Eq. (\ref{move_model}) have the same expression as Eq. (\ref{Model}), which is required for Galilean invariance. Hence, the variation in the magnitude of the velocity, $c$, will not influence the process of SSB. However, in the current system, Eq. (\ref{move_model}) does not have the same expression as Eq. (\ref{Model}), and the magnitude of the velocity, $c$, will definitely influence the process of the SSB.
It is necessary to mention that Galilean invariance can also be broken by applying a periodic boundary condition to the system. With a periodical boundary condition, the moving velocity changes to a rotational angular velocity. The SSB in the DWP system with a toroidal trap depends on the magnitude of the rotational velocity \cite{Guihua2017}. A recent study found that such a rotating system can emulate the SO coupling system \cite{Haohuang2019Opt}, which reveals the internal connection between the rotating system and the effect of SO coupling.

The SSB of the soliton can be obtained by solving the stationary solutions to Eqs. (\ref{move_model}). The definition of $\delta$ in Eq. (\ref{delta}) remains valid by replacing $x'$ with $x$. A Numerical simulation shows that increasing the values of $c$ can lead to SSB of the soliton. The reason the SSB is induced by $c$ can be explained by Eq. (\ref{move_model}). The last term in Eq. (\ref{move_model}) creates a linear mixing between the two components. The linear mixing has the same profile of $\lambda(x')$, which enhances the difference between the local maximum (at $x'=0$) and the two adjacent minima (at $x'=\pm x_{0}$). The value of $c$ is the strength of such linear mixing. Hence, increasing the value of $c$ may reduce the coupling between the two wells, which makes the SSB easier to induce.

By selecting different values of $x_{0}$, displacement SSB and bimodal SSB can be found and adjusted by the parameter $c$. Fig. \ref{movingbifuraction}a shows a typical example of the bifurcation map of the bimodal SSB induced by $c$. Typical examples of the symmetric and asymmetric solitons, which are labeled by `a', `b', and `c' in Fig. \ref{movingbifuraction}a, are displayed in Fig. \ref{moving}. It is interesting to note that the evolution of the unstable symmetric solution, which is shown in Fig. \ref{moving}(c3), shows a typical Josephson Oscillation \cite{Albiez2005}. Josephson Oscillation in a DWP with homogeneous SO coupling and a moving system were report in Refs. \cite{Zhang2012,Haoxu2014}, respectively. This result implies that the current system may have potential in matter-wave interferometry \cite{Shin2004,Schumm2005}.

Fig. \ref{movingbifuraction}b displays $g_{\mathrm{cr}}$ as a function of $c$ with different values of $x_{0}$. The figure shows that an increase in $c$ causes a transition between displacement SSB and bimodal SSB. The dividing point between displacement SSB and bimodal SSB can be adjusted by the values of $x_{0}$ and $c$.

\section{Conclusion}

The objective of this work was to study the SSB of a soliton created in spinor BECs with spatially confined SO coupling formed by a two-spot laser beam. Two types of SSB, displacement symmetry breaking and bimodal symmetry breaking, are found in the system. Both types are supercritical, which indicates that these two kinds of SSB are tantamount to a phase transition of the second kind. By increasing the strength of the nonlinearity, displacement SSB tends to occur when the two spots of the SO coupling are close enough; however, when the distance between the two spots becomes large, bimodal SSB will occur, taking the place of displacement SSB. The explanation for the transition between these two kinds of SSB is discussed in detail in this paper. SSB in the moving reference is also considered because the SO coupling system in the moving reference frame is a nontrivial issue. The results show that the velocity plays an important role in influencing the SSB. Increasing the velocity may not only help to induce the SSB but also help the transition from displacement SSB to bimodal SSB.

A natural continuation of the current work is to consider this problem in 2D, assuming the BECs are trapped in a 2D plane and illuminated by two SO coupling spots. Hence, the inclusion of 2D vortices may generate more degrees of freedoms to influence the SSB. A more challenging option is to consider the current setup in a full 3D geometry.

\begin{acknowledgments}
This work was supported by NNSFC (China) through grants No. 11874112,
11575063, Foundation for Distinguished Young Talents in Higher Education of Guangdong No. 2018KQNCX279, 2018KQNCX009, and the Special Funds for the Cultivation of Guangdong College Students Scientific and Technological Innovation No. pdjh2019b0514.
\end{acknowledgments}

\textbf{Conflict of interest} The authors declare that there is no conflict
of interest to report.

\end{document}